\documentclass[reprint,superscriptaddress,amsmath,amssymb,aps,prl]{revtex4-2}

\usepackage{color}
\usepackage{graphicx}
\usepackage{dcolumn}
\usepackage{bm}

\usepackage{amsmath}
\usepackage{tikz}
\usetikzlibrary{matrix,fit}

\usepackage[T1]{fontenc}
\usepackage{newtxtext}
\usepackage{newtxmath}
\usepackage{upgreek}
\usepackage{mhchem}
\usepackage{hyperref}
\hypersetup{colorlinks=true,linkcolor=blue,citecolor=blue,urlcolor=blue}

\begin{document}

\preprint{APS/123-QED}

\title{Pressure Tuning of Layer-hybridized Excitons in Trilayer WSe$_2$}

\author{Xuan Zhao}
\affiliation{Beijing National Laboratory for Condensed Matter Physics, Institute of Physics, Chinese Academy of Sciences, Beijing, 100190, China}
\affiliation{School of Physical Sciences, University of Chinese Academy of Sciences, Beijing 100049, China}

\author{Jing Song}
\email{jingsong@iphy.ac.cn}
\affiliation{Beijing National Laboratory for Condensed Matter Physics, Institute of Physics, Chinese Academy of Sciences, Beijing, 100190, China}

\author{Wenqi Xiong}
\affiliation{Institute of Quantum Materials and Physics, Henan Academy of Sciences, Zhengzhou, 450046, China}

\author{Qianying Hu}
\affiliation{Beijing National Laboratory for Condensed Matter Physics, Institute of Physics, Chinese Academy of Sciences, Beijing, 100190, China}
\affiliation{School of Physical Sciences, University of Chinese Academy of Sciences, Beijing 100049, China}

\author{Yuxuan Song}
\affiliation{Beijing National Laboratory for Condensed Matter Physics, Institute of Physics, Chinese Academy of Sciences, Beijing, 100190, China}
\affiliation{School of Physical Sciences, University of Chinese Academy of Sciences, Beijing 100049, China}

\author{Xin He}
\affiliation{Beijing National Laboratory for Condensed Matter Physics, Institute of Physics, Chinese Academy of Sciences, Beijing, 100190, China}
\affiliation{School of Physical Sciences, University of Chinese Academy of Sciences, Beijing 100049, China}

\author{Tianzhong Yang}
\affiliation{Beijing National Laboratory for Condensed Matter Physics, Institute of Physics, Chinese Academy of Sciences, Beijing, 100190, China}

\author{Song Liu}
\affiliation{Institute of Microelectronics, Chinese Academy of Sciences, Beijing 100029, China}

\author{Shengjun Yuan}
\affiliation{Key Laboratory of Artificial Micro- and Nano-structures of Ministry of Education and School of Physics and Technology, Wuhan University, Wuhan 430072, China}
\affiliation{Wuhan Institute of Quantum Technology, Wuhan, 430206, China}

\author{Hongyi Yu}
\affiliation{ Guangdong Provincial Key Laboratory of Quantum Metrology and Sensing \& School of Physics and Astronomy, Sun Yat-Sen University (Zhuhai Campus), Zhuhai 519082, China}
\affiliation{State Key Laboratory of Optoelectronic Materials and Technologies, Sun Yat-Sen University (Guangzhou Campus), Guangzhou 510275, China}

\author{Yang Xu}
\email{yang.xu@iphy.ac.cn}
\affiliation{Beijing National Laboratory for Condensed Matter Physics, Institute of Physics, Chinese Academy of Sciences, Beijing, 100190, China}
\affiliation{School of Physical Sciences, University of Chinese Academy of Sciences, Beijing 100049, China}


\begin{abstract}
We demonstrate dynamic pressure tuning (0-6.6 GPa) of layer-hybridized excitons in AB-stacked trilayer WSe$_2$ via diamond-anvil-cell-integrated reflectance spectroscopy. Pressure-controlled interlayer coupling manifests in enhanced energy-level anti-crossings and oscillator strength redistribution, with Stark shift analysis revealing a characteristic dipole moment reduction of 11\%. Notably, the hybridization strength between the intra- and interlayer excitons triples from $\sim$10 meV to above $\sim$30 meV, exhibiting a near-linear scaling of 3.5$\pm$0.2 meV/GPa. Spectral density simulations resolve four distinct components, i.e., intralayer ground/excited and interlayer ground/excited excitons, with their relative weights transitioning from one component dominant to strongly hybridized at higher pressures. Our findings highlight the potential for controlling excitonic properties and engineering novel optoelectronic devices through interlayer compression.
\end{abstract}


\maketitle

Two-dimensional (2D) vdW heterostructures assembled from atomically thin semiconducting transition metal dichalcogenides (TMDCs) exhibit unprecedented physical properties arising from a unique combination of the spin and valley degrees of freedom and show vast potentials for fascinating optoelectronic and valleytronic applications \cite{CIARROCCHI2022Excitonic,SIERRA2021Van,RIVERA2018Interlayer,WANG2018Colloquium,XIAO2012Coupled,GONG2013Magnetoelectric,JONES2014Spin,MAK2012Control,ZENG2012Valley,CAO2012Valleyselective,WU2013Electrical}. The spin-valley locking effect, originally known to be dominant in TMDC monolayers, also plays a crucial role in the multilayers. For example, a new type of every-other-layer dipolar excitons in the pristine 2H-trilayer WSe$_2$ and MoSe$_2$ has recently been identified \cite{ZHANG2023,DU2024New,FENG2024Highly}. The dipole moment and Stark shift of such every-other-layer excitons, with a middle layer acting as a barrier, are twice that of interlayer excitons in TMDC homo- or hetero- bilayers. Furthermore, due to their increased stability and extended lifetime arising from the large spatial separation, they emerge as promising candidates for achieving giant optical nonlinearity and exciton condensation \cite{SUN2024Dipolar,COMBESCOT2017Bose,DEBNATH2017Exciton,FOGLER2014Hightemperature,GU2024Giant}.

Engineering the interlayer coupling lies at the core of exploring various 2D vdW heterostructures. Notable examples include proximity-induced spin-orbit coupling \cite{ISLAND2019Spin} and magnetism \cite{ZHONG2020Layerresolved}, emergent phenomena such as correlated insulating states \cite{XU2020Correlated} and unconventional superconductivity \cite{CAO2018Unconventional} in moiré superlattices, and new functionalities such as spontaneous photovoltaic effect \cite{AKAMATSU2021Van} and wide-range photodetectors \cite{LEE2012MoS2}. However, accurately determining and manipulating the interlayer coupling remains an outstanding challenge. In recent years, hydrostatic pressure has emerged as a powerful knob for adjusting the vdW gap and introducing novel structural and electronic modifications \cite{ZHANG2022Pressure,PIMENTAMARTINS2023Pressure,DOU2016Probing,YE2016Pressureinduced,CHEN2017Pressurizing,XIA2021Strong,SONG2019Switching,QIAN2022Unconventional,MA2021Robust,ZHAO2021Dynamic,LI2022Engineering,KE2019Large,LI2019Pressurecontrolled,FAN2015Electronic,YANKOWITZ2019Tuning,YANKOWITZ2018Dynamic,HUANG2021Layerdependent}. Previous pressure studies on vdW materials have often lacked the integration of high-quality samples, gate-tunability, low-temperature conditions, and reflectance measurements -- key factors essential for numerous important experimental discoveries. In this work, we developed a cryogenically compatible diamond anvil cell (DAC) featuring both in-situ electrical and optical access for high-quality vdW heterostructure samples. Trilayer WSe$_2$ hosting an intriguing interplay between intra- and interlayer excitons, has been chosen here for studying the effects of hydrostatic pressure up to 6.6 GPa. The setup allows us to systematically investigate the spectra evolution of the excitonic resonances as the layer spacing decreases, offering crucial insights into the layer hybridization process.

Figure 1a shows a schematic of our experimental set-up. Pressure ($P$) is applied uniformly on the sample by compressing the DAC chamber sealed with liquid pressure-transmitting medium (PTM), and further calibrated by the energy shift of the ruby $R$1 line \cite{PIMENTAMARTINS2023Pressure,YANKOWITZ2019Tuning,YANKOWITZ2018Dynamic,HUANG2021Layerdependent,ZHANG20202D,ZHAO2015Pressure}(see Supplemental Materials \cite{supple} for DAC details). Pre-patterned Ti/Au electrodes along with gold wires enable electrical access, while the insulating layer between the gasket and the diamond culet prevents short contacts. To independently tune the vertical electric field ($E_{\mathrm{z}}$) and doping density, we fabricate dual-gate devices as illustrated in Fig. 1b. The optically-active trilayer WSe$_2$ is fully encapsulated between few-layer graphite and hBN, which serve as local gate electrodes and dielectrics on both sides. The encapsulation also helps maintain ultra-clean interfaces and ensure the high optical quality in the whole pressure range. Exciton hybridization behaviors are primarily studied using optical reflectance micro-spectroscopy with a probing spot of approximately 2 $\mathrm{\mu}$m. The entire DAC is maintained at a cryogenic temperature of 8 K during measurements.

\begin{figure}
	\includegraphics[width=250pt]{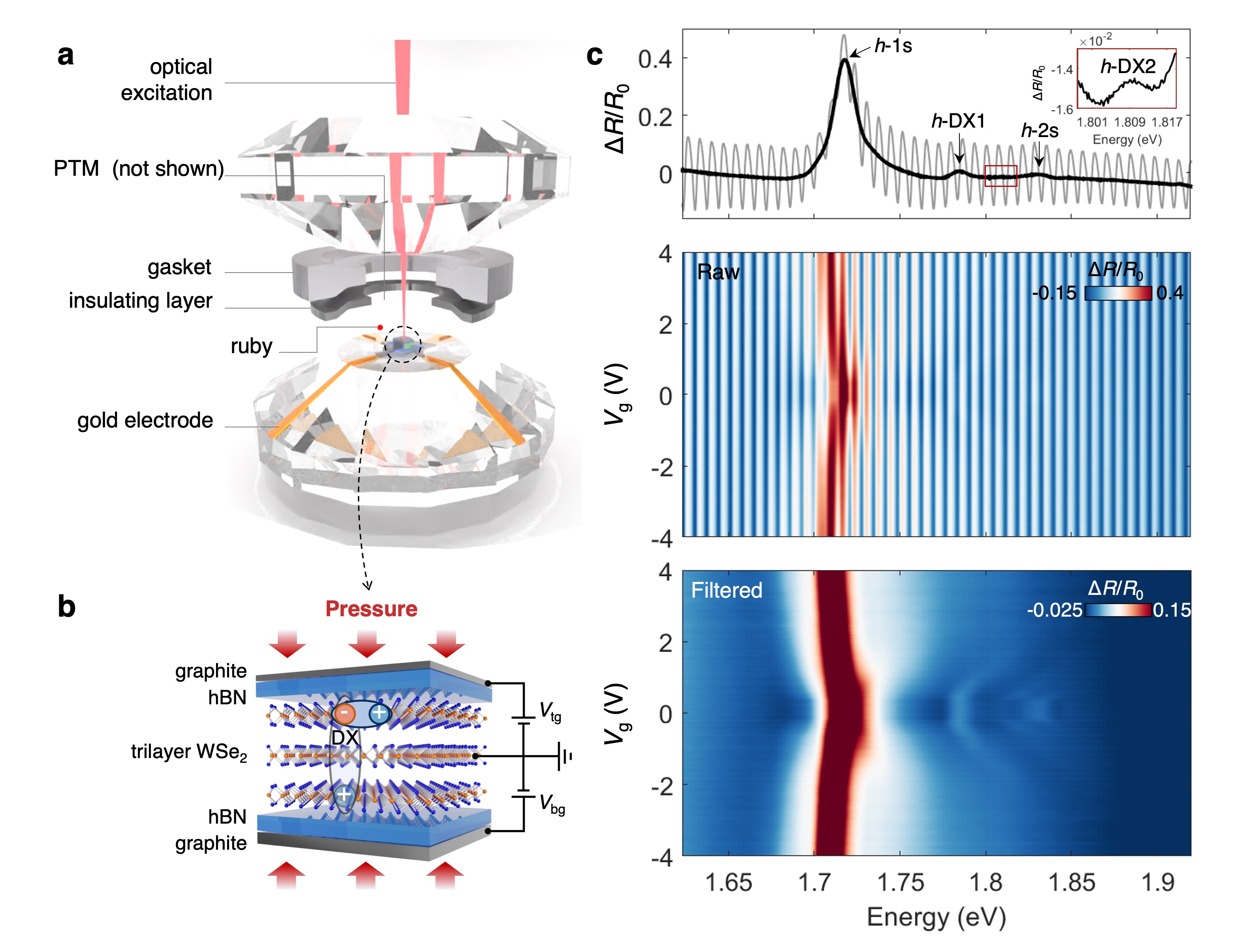}
	\caption{\label{fig:1} (a) Illustration of the diamond anvil cell (DAC) set-up. Components of the DAC are displayed separately for clarity. The sample is attached to the diamond culet and in contact with the pre-patterned gold electrodes, allowing the electrical access. Mineral oil sealed in the DAC sample chamber acts as PTM for uniform compression on the sample. The DAC is maintained at a temperature of 8 K during measurements.  (b) Schematic of the tri-layer WSe$_{2}$ device under pressure. Dual-gate configuration allows an independent tuning of the doping and electric field.  (c) Reflectance contrast spectra in the DAC at 0 GPa (without PTM). Upper panel shows data at zero gate voltage. Interference between two diamond culets induce severe oscillations, which can be filtered out by fast Fourier transformation (FFT). The peaks located at 1.718, 1.785, and 1.830 eV are attributed to $h$-1s, $h$-DX, and $h$-2s states, respectively. Middle and lower panels are the doping dependencies of raw and filtered spectra, respectively.}
\end{figure}

To examine the exciton resonances in trilayer WSe$_2$, we first present the reflectance contrast ($\Delta R/R_0$) measurements performed in the DAC without filling the PTM ($P = 0$ GPa). The upper panel of Fig. 1c displays the $\Delta R/R_0$ spectra at charge neutrality and zero electric field. The interference between the two parallel diamond culets creates severe oscillations in the spectra, which can be removed using the fast Fourier transformation (FFT) filtering (see Supplementary Fig. 3 \cite{supple}). Raw and FFT-filtered data are displayed in gray and black, respectively.

Peaks observed at 1.718, 1.785, and 1.830 eV are respectively attributed to $h$-1s, $h$-DX1, and $h$-2s states (the prefix “$h$-” denotes hybridization \cite{ZHANG2023} and will be discussed later). The 1s and 2s stand for the intralayer excitonic ground state and the first excited state at the K/K' valley, respectively. The 2H-stacked trilayer WSe$_2$ inherently features adjacent layers rotated by 180 degrees, resulting in alternately arranged spin-valley-locked K and K' valleys and giving rise to the initially dark every-other-layer exciton DX1 \cite{ZHANG2023,FENG2024Highly,GU2024Giant}. The electron and hole constituents of DX1 predominantly segregate into the next-nearest layers, as illustrated in Fig. 1b. It gains finite oscillator strength through hybridization with intralayer excitons and hence denoted as $h$-DX1 (schematically shown in Fig. 1b). Additionally, we notice a peak with relatively weak intensity at 1.810 eV, which has not been documented before and is identified as the interlayer hybridized excited exciton state $h$-DX2. Middle and lower panels show the raw and FFT-filtered doping-dependent reflectance contrast spectra (maintaining $E_{\mathrm{z}}=0$). The filtered spectra clearly reveal fine features of trions or exciton-polarons upon electron/hole doping, consistent with the previous report\cite{ZHANG2023}.

To systematically study the exciton hybridization effect, we have measured the reflectance contrast spectra of the trilayer WSe$_2$ devices at 10 different pressures over the range of $0-6.6$ GPa. The electric-field-dependent spectra at charge neutrality and four representative pressures are shown in Fig. 2a (more data shown in Supplementary Fig. 5 \cite{supple}). Due to the continuously increasing bandgap with pressure, the energy is plotted in different ranges for clarity. We first use the spectra at ambient pressure (left panel of Fig. 2a) as an example to illustrate the effect of electric fields.

\begin{figure*}
	\includegraphics[width=480pt]{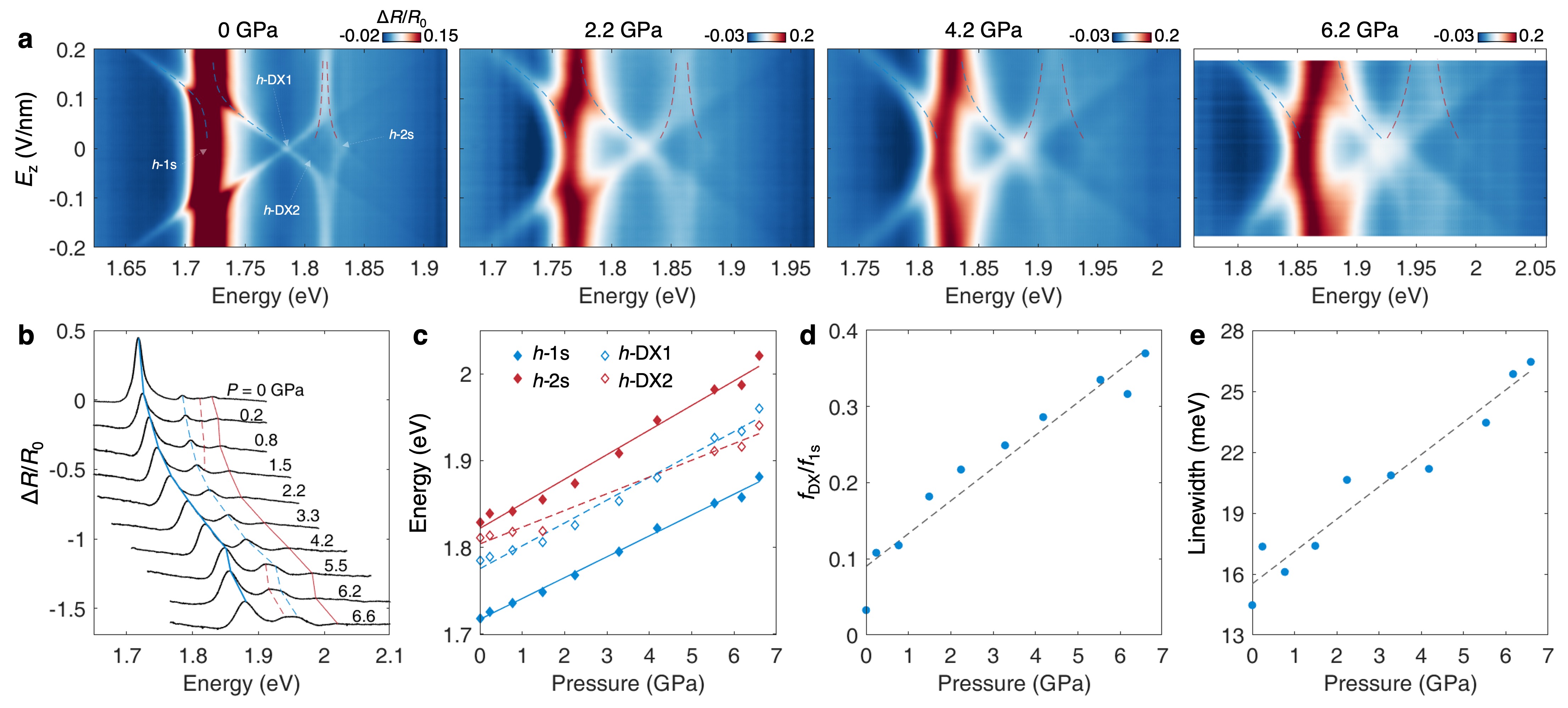}
	\caption{\label{fig:2} (a) Electric-field-dependent reflectance contrast spectra at 4 representative pressures. The anti-crossing (guided by dashed curves) between the $h$-1s and $h$-DX states grows with increasing the pressure, and the $h$-2s state at finite electric fields splits into two branches. The intensity of the $h$-DXs state relative to the $h$-1s state gets enhanced with increasing pressure. (b) Spectral linecuts (vertically shifted for clarity) at charge neutrality and zero electric field at pressures of $0-6.6$ GPa. The solid blue, dashed blue, dashed red, and solid red lines trace the $h$-1s, $h$-DX1, $h$-DX2, and $h$-2s exciton peaks, respectively. (c) Peak energies extracted from Lorentzian fitting of the spectra in (b), showing linear blueshifts at rates of $19-28$ meV per GPa. (d) The oscillator strength of dark excitons $f_{\mathrm{DX}}$ (sum of DX1 and DX2) relative to that of the 1s state $f_{\mathrm{1s}}$ as a function of the pressure. (e) The full-width at half-maximum (FWHM) of the $h$-1s state continuously increases with the pressure.}
\end{figure*}

The X-shaped feature centering around 1.785 eV ($h$-DX1) stems from the quantum-confined Stark shift of every-other-layer excitons with opposite dipole orientations. The slope of the “X” feature directly gives the effective dipole moment (defined as
\begin{equation}
    \mu_\mathrm{eff}(P)=\frac{\varepsilon _{\mathrm {hBN}}(P)}{\varepsilon _{\mathrm {WSe}_{2}}(P)}\cdot 2ed
\end{equation}
, where $\varepsilon _{\mathrm {hBN}}$ and $\varepsilon _{\mathrm {WSe}_{2}}$ are the dielectric constants of hBN and WSe$_2$, $e$ is the elementary charge, $d$ is the interlayer distance of WSe$_2$, and $2$ signifies the double spacing, see Supplemental Materials \cite{supple}). We obtain $\mu_{\mathrm{eff}} = 0.62$ $e \mathrm{nm}$ at $P = 0$ GPa, which agrees well with the previous results\cite{ZHANG2023,GU2024Giant}. Another prominent feature is the avoid-crossing behavior between the every-other-layer exciton branch (DX1) and the intralayer 1s exciton branch at $\sim$1.72 eV (indicated by the blue dashed curves). This happens when the two excitons are tuned into resonance and there is finite carrier hopping between them. The anti-crossing gap equals twice the hybridization strength.

Additionally, we notice another two faint “X” features at 1.810 eV and 1.830 eV, respectively. They both correspond to the Stark shifts of excited exciton states (corresponding to $h$-DX2 and $h$-2s in Fig. 1 respectively), with one branch growing linearly with the electric field and the other branch (highlighted by the dashed red curves) evolving into an electric-field-insensitive exciton state (identified as the intralayer 2s state). The linear growing branch shares the same dipole moment $\mu_{\mathrm{eff}}$ as DX1, exhibiting characteristic of every-other-layer excitons. All these features are puzzling at first sight and cannot be fully understood without the pressure tuning.

As pressure increases, the exciton hybridization is enhanced, manifesting through several spectral changes. First, the anti-crossing gap between the lower-energy branch of the interlayer exciton (DX1) and the intralayer exciton (1s) increases under pressure (guided by the blue dashed curves), accompanied by an intensity growth of $h$-DX1. Second, the excited exciton state at finite $E_{\mathrm{z}}$ splits into two clearer branches, displaying a pair of mirror-symmetrical curves with increasing separation (guided by the two red dashed curves) at higher pressures.

We further carefully examine the spectral linecuts at zero electric field (Fig. 2b). As guided by the solid and dashed blue (red) curves for $h$-1s and $h$-DX1 ($h$-2s and $h$-DX2), all excitonic resonances experience blue shifts under pressure, indicating a continuous growth of optical bandgap at the K/K' valleys. Figure 2c shows the linear increase of the hybridized exciton resonance energies extracted via Lorentzian fitting. There is an energy crossover between $h$-DX1 and $h$-DX2 at a pressure of $\sim$4 GPa. The $h$-1s, $h$-DX1, and $h$-2s state (denoted as solid blue, hollow blue, and solid red diamonds) exhibit blueshifts at rates of 24.0$\pm$0.6, 26.3$\pm$1.1, and 28.3$\pm$1.3 meV/GPa, respectively, whereas the $h$-DX2 (denoted as hollow red diamonds) has a lower rate of 19.3$\pm$1.3 meV/GPa. The rate of $h$-1s is close to the value reported for similar systems (23.9 meV/GPa for trilayer MoS$_2$ \cite{DOU2016Probing} and 27 meV/GPa for bilayer WSe$_2$ \cite{YE2016Pressureinduced}). However, we note that these excitonic resonances are strongly mixed states. Their evolution cannot be simply understood without decomposing them into the initial states before hybridization. Moreover, the oscillator strength of $h$-DXs (sum of $h$-DX1 and $h$-DX2) compared to $h$-1s continuously increases (Fig. 2d), signifying the increase of hybridization strength between intra- and interlayer excitons. The linewidth of $h$-1s continuously broadens (Fig. 2e), probably stemming from the increased strain gradient in the sample.

\begin{figure*}
	\includegraphics[width=480pt]{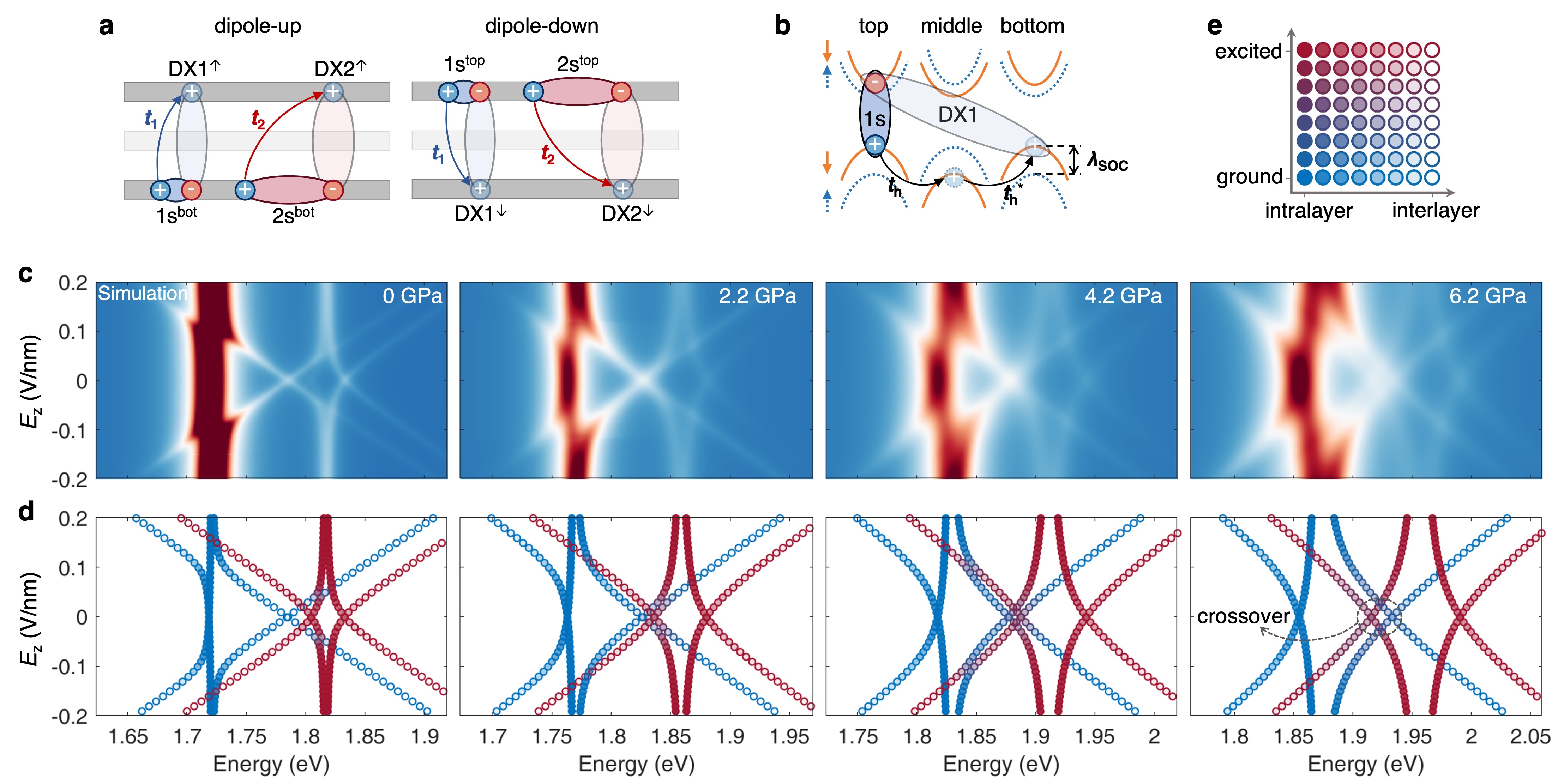}
	\caption{\label{fig:3} (a) Schematic of the modified exciton hybridization model, wherein the interlayer excited state (DX2) is introduced, and the 1s and 2s excitons in the top (bottom) WSe$_2$ layer exclusively couple to the dipole-down (dipole-up) interlayer excitons. (b) Schematic of the band structure of the 2H-stacked tri-layer WSe$_2$. Alternately arranged spin-valley-locked K and K’ valleys result in every-other-layer excitons (DXs, denoted by the translucent oval), which gain oscillator strength through hybridization (dominant by the hole tunneling process) with the intralayer bright excitons (denoted by the blue oval). (c) Simulations of the electric-field-dependent spectral density at pressures in accordance with Fig. 2a, exhibiting excellent agreement with the experimental data. (d) Resonance energies and bare state component of the hybridized excitons. Interestingly, the hybridized 2s state ($h$-DX2) gradually crosses over to the low energy side of the hybridized 1s state ($h$-DX1) at higher pressures as marked by the dashed circle. (e) Legend for (d). The intralayer and interlayer species are strongly mixed under pressure. Different orbitals are represented by a mix of colors. The solid blue and red circles denote the pure 1s and 2s states. The transparency of the circle's interior symbolizes the intra-/inter-layer characteristics (definitions provided in the Supplemental Materials \cite{supple}).}
\end{figure*}

The excitonic hybridization model proposed in the previous study \cite{ZHANG2023} does not incorporate the interlayer excited excitons and fails to acknowledge that the carrier tunneling between every other layers essentially does not occur for conduction-band electrons. To describe the pressure-tuned exciton hybridization in trilayer WSe$_2$ and gain a deeper microscopic understanding, we employ a modified Hamiltonian written as


\[
H = 
\begin{tikzpicture}[baseline=(m.center)]
\matrix (m) [matrix of math nodes, 
             left delimiter=(, right delimiter=), 
             row sep=0.2em, 
             column sep=0.2em, 
             inner sep=0.1em] {
  e_{1\mathrm s}^{\mathrm{bot}} & t_1 & 0 & t_6 & 0 & 0 & 0 & 0\\
  t_1 & e_{\mathrm{DX1}}^{\uparrow} & t_3 & 0 & 0 & 0 & 0 & 0\\
  0 & t_3 & e_{2\mathrm s}^{\mathrm{bot}} & t_2 & 0 & 0 & 0 & 0\\
  t_6 & 0 & t_2 & e_{\mathrm{DX2}}^{\uparrow} & 0 & 0 & 0 & 0\\
  0 & 0 & 0 & 0 & e_{1\mathrm s}^{\mathrm{top}} & t_1 & 0 & t_6\\
  0 & 0 & 0 & 0 & t_1 & e_{\mathrm{DX1}}^{\downarrow} & t_3 & 0\\
  0 & 0 & 0 & 0 & 0 & t_3 & e_{2\mathrm s}^{\mathrm{top}} & t_2\\
  0 & 0 & 0 & 0 & t_6 & 0 & t_2 & e_{\mathrm{DX2}}^{\downarrow}\\
};

\node [fit=(m-1-1) (m-4-4), draw, dashed, inner sep=0pt] {};

\node [fit=(m-5-5) (m-8-8), draw, dashed, inner sep=0pt] {};
\end{tikzpicture}
\]
, where each of the eight basis represents the energy of a non-hybridized bare exciton. They are intralayer ground and excited states in the top and bottom layer (denoted as $\left | \mathrm{1s}^{\mathrm{top}} \right \rangle$, $\left | \mathrm{1s}^{\mathrm{bot}} \right \rangle$, $\left | \mathrm{2s}^{\mathrm{top}} \right \rangle$, $\left | \mathrm{2s}^{\mathrm{bot}} \right \rangle$) and the interlayer ground and excited states with dipole orientations up and down (denoted as $\left | \mathrm{DX1}^{\uparrow} \right \rangle$, $\left | \mathrm{DX1}^{\downarrow} \right \rangle$, $\left | \mathrm{DX2}^{\uparrow} \right \rangle$, $\left | \mathrm{DX2}^{\downarrow} \right \rangle$, where “$\uparrow$” or “$\downarrow$” indicates the dipole orientation), as illustrated in Fig. 3a. The energies of interlayer excitons are $e_{\mathrm{DX1}(\mathrm{DX2})}^{\uparrow}=e_{\mathrm{DX1}(\mathrm{DX2})}-E_\mathrm{z}\times \mu_\mathrm{eff}$ and $e_{\mathrm{DX1}(\mathrm{DX2})}^{\downarrow}=e_{\mathrm{DX1}(\mathrm{DX2})}+E_\mathrm{z}\times \mu_\mathrm{eff}$, with $\mu_\mathrm{eff}E_\mathrm{z}$ denoting the Stark shift (assuming to be the same for DX1 and DX2 in the fitting). Analysis of the orbital composition of the band edge at K/K’ point leads to the conclusion that the interlayer electron hopping is forbidden by the $\mathrm{C}_3$ symmetry (see Supplemental Materials \cite{supple} for details). The hybridization strengths between intra- and interlayer excitons are dominated by a second-order hole-tunneling process between the top and bottom layers with $t = t_h^2/ \lambda_{\mathrm{SOC}}$, where $\lambda_{\mathrm{SOC}} \sim$450 meV \cite{GONG2013Magnetoelectric} is the valence band spin-orbit splitting and th is the hole hopping amplitude between adjacent layers (see Fig. 3b). Hence the bottom (top) layer excitons only couple to the dipole-up (dipole-down) every-other-layer excitons and are arranged in the upper-left (lower-right) block of the Hamiltonian. As depicted in Fig. 3a, the off-diagonal term $t_1$ ($t_2$) represents the hybridization strength between 1s and DX1 (2s and DX2), while $t_3$ ($t_6$) corresponds to the minor hybridization effect between 2s and DX1 (1s and DX2). Diagonalizing the Hamiltonian would give four degenerate eigenstates (i.e., hybridized excitons $h$-1s, $h$-DX1, $h$-DX2, and $h$-2s, respectively) at zero electric field ($E_{\mathrm{z}}=0$) and eight eigenstates at finite electric fields.
	
Based on this model, Fig. 3c-d displays the spectral density simulation and the eigenvalues of hybridized excitons at the same pressures as in Fig. 2a (comparison of experiments and simulations at more pressures are shown in Supplementary Fig. 5 \cite{supple}). A Lorentzian lineshape is used in the spectral density calculation according to the microscopic analysis of excitons in few-layer MoSe$_2$ \cite{SPONFELDNER2022Capacitively}. The experimental and calculated spectra exhibit excellent consistency, further verifying our model. With the extracted eigenvalues, it can be seen more clearly that, the anti-crossing gap is significantly enhanced at higher pressures. There are four fundamental exciton species, i.e. the interlayer/intralayer ground/excited states. Figure 3d also visualizes the fundamental exciton constitution in a manner illustrated in Fig. 3e. The full colors of blue and red represent pure ground and excited states, and the empty/filled circles correspond to the interlayer/intralayer excitons. The exciton states under pressure are highly hybridized from these four components, as indicated by the mixing of blue/red colors and the semi-transparent interior of the circle markers. As the hybridization strength increases with pressure, the $h$-DX2 states become energetically lower than $h$-DX1 states near zero $E_{\mathrm{z}}$ at 6.2 GPa, as marked by the dashed circle in Fig. 3d.

Figure 4 depicts parameters extracted from the model fitting (see more in Supplementary Fig. 6 \cite{supple}). The pressure-dependent un-hybridized exciton energies ($e_{\mathrm{1s}}$, $e_{\mathrm{2s}}$, $e_{\mathrm{DX1}}$, and $e_{\mathrm{DX2}}$) show continuous blueshifts (Fig. 4a). This method reveals the energies of the four bare excitons tuned by pressure, which are otherwise difficult to deconvolve from the intricate energy splitting caused by hybridization effects. The energy of the intralayer 1s state ($e_{\mathrm{1s}}$) exhibits a linear shift about 26.5 ± 0.8 meV/GPa (guided by the dashed line in Fig. 4a), while the other exciton species shift at lower rates. Intriguingly, the energies of DX2 and 2s are nearly degenerate throughout the measured pressure range, resulting in the obvious energy splitting (into $h$-DX2 and $h$-2s) at $E_{\mathrm{z}}=0$. This also explains the nearly symmetrical spectral behavior of $h$-2s and $h$-DX2 (see red dashed curves in Fig. 2a and red symbols in Fig. 3d) under various $E_{\mathrm{z}}$ values. The energy separation between the 1s and 2s excitons decreases from $\sim$96.4 meV to $\sim$81.0 meV at 6.6 GPa (see Supplementary Fig. 7 \cite{supple}), indicating a suppression of intralayer exciton binding energy likely due to an increase in the dielectric screening \cite{YANKOWITZ2018Dynamic} with pressure. Conversely, there is an increase of energy separation between DX1 and DX2, indicating an enhanced interlayer exciton binding energy, predominantly influenced by the reduced electron-hole spatial separation (see the change of $\mu_{\mathrm{eff}}$ with pressure in Fig. 4b). Specifically, there is an 11\% reduction in $\mu_{\mathrm{eff}}$ at 6.6 GPa compared to its value at ambient pressure, reflecting the sensitivity of interlayer distance of WSe$_2$ to the out-of-plane compression (Supplementary Fig. 8 \cite{supple}).

\begin{figure}
	\includegraphics[width=250pt]{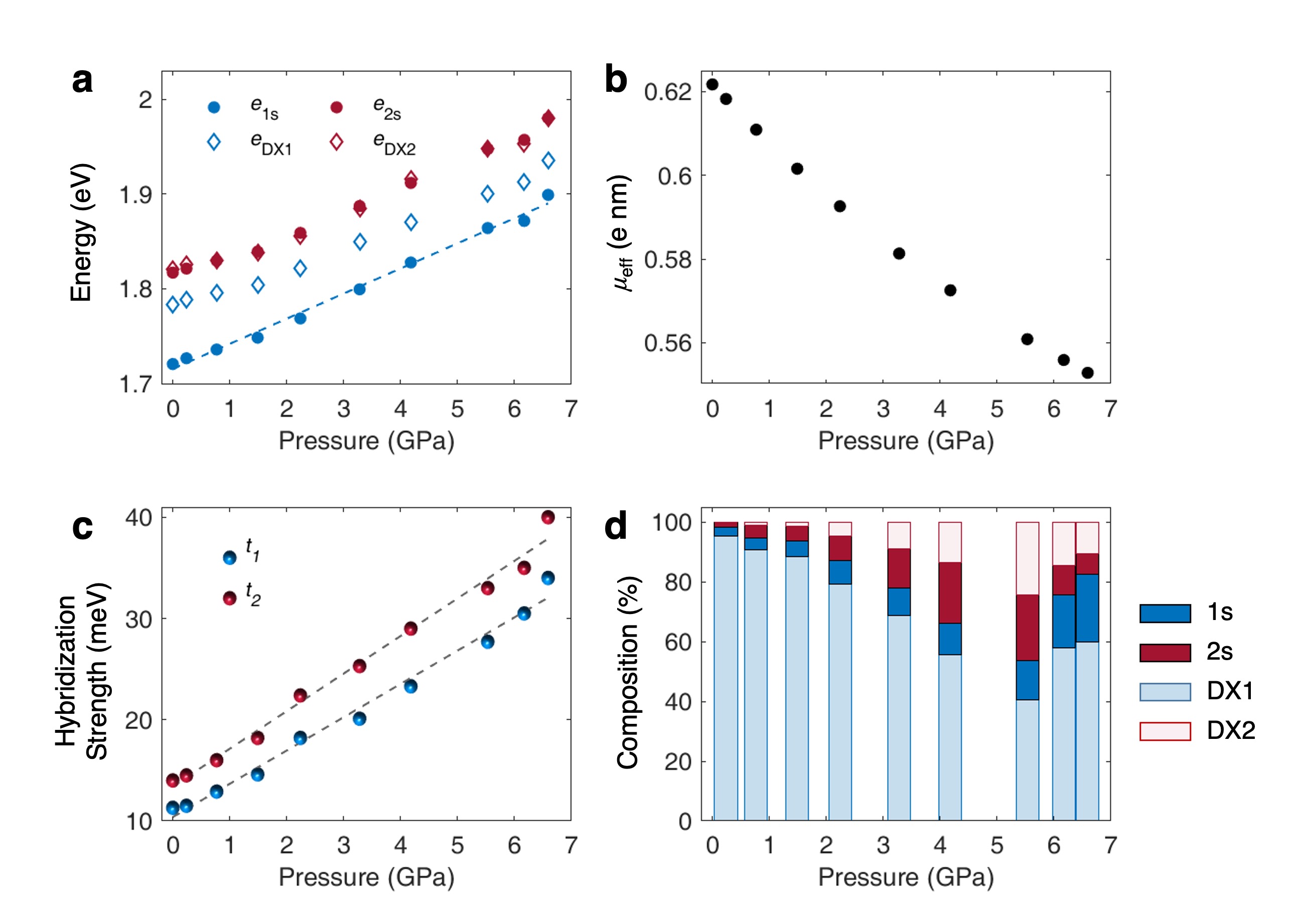}
	\caption{\label{fig:4} (a) Pressure-dependent bare (un-hybridized) exciton energies, which exhibit continuous blue shifts. (b) The effective dipole moment of the every-other-layer exciton (defined in the main text) decreases with pressure, showing an 11\% reduction at 6.6 GPa compared to the value at ambient pressure. (c) Pressure-dependent hybridization strength $t_1$ and $t_2$, both of which increase at a similar rate under pressures up to 6.6 GPa (3.3 and 3.7 meV/GPa, respectively). (d) Composition of $h$-DX1.}
\end{figure}

Hybridization strength between 1s and DX1 ($t_1$) is extracted to be 11.3 meV (14 meV for the hybridization $t_2$ between 2s and DX2) at ambient pressure, consistent with 10$\pm$2 meV reported by ref.  \cite{GU2024Giant} and double the value in ref.  \cite{ZHANG2023}. The hybridization strength is hence effectively tuned by pressure via reducing the interlayer spacing. As shown in Fig. 4c, the increasing rate of hybridization strength of 1s-DX1 ($t_1$) and 2s-DX2 ($t_2$) are similar (3.3 and 3.7 meV/GPa, respectively). They reach values of $\sim$34 meV and $\sim$40 meV at 6.6 GPa, respectively. The slightly larger values of $t_2$ probably stem from the larger spatial extension of the excited exciton states. Using $t_1$ for the estimation of $ t=t_h^2/\lambda_{\mathrm{SOC}}$, hole hopping $t_h$ between adjacent layers is extracted to be $\sim$71 meV at ambient pressure, agreeing with the value of 67 meV reported previously \cite{GONG2013Magnetoelectric}. The amplitude of th as a function of pressure is shown in Supplementary Fig. 6 \cite{supple}, increasing at a rate of 7.7$\pm$0.2 meV/GPa.

The component of each hybridized exciton is defined by the relevant elements of the transformation matrix $U$ (see Supplemental Materials \cite{supple}). The mixing between intralayer and interlayer exciton states becomes significant under out-of-plane compression for both $h$-DX1 and $h$-DX2, accounting for their enhanced oscillator strengths under pressure. In Fig. 4d, we show the four decomposed excitonic elements ($\left | \mathrm{1s} \right \rangle$, $\left | \mathrm{2s} \right \rangle$, $\left | \mathrm{DX1} \right \rangle$, and $\left | \mathrm{DX2} \right \rangle$) that comprise $h$-DX1. The intralayer component increases from 4.7\% at ambient pressure to 35.1\% at 5.5 GPa. All four components contribute significantly at high pressures. Composition of other hybridized excitons under different pressures can be found in Supplementary Fig. 9 \cite{supple}. 

In summary, we have implemented hydrostatic pressure to dynamically control the vdW gap and induce prominent intra- and inter-layer exciton hybridization in high-quality trilayer WSe$_2$ devices. Such strongly layer-hybridized excitonic states share the characteristics of both the intra- and inter-layer excitons, i.e., large oscillator strength, long lifetime, and large out-of-plane dipoles. This would collectively facilitate the study for achieving giant optical nonlinearity and exciton condensation \cite{SUN2024Dipolar,COMBESCOT2017Bose,DEBNATH2017Exciton,FOGLER2014Hightemperature,GU2024Giant}. By clarifying the dominant second-order hole tunneling process, we extract the evolution of hole hopping amplitude between adjacent layers under pressure. Our results can advance the comprehension of interlayer coupling in TMDC and demonstrate an experimental approach for further exploring the pressure-adjusted correlation and topological effects in vdW heterostructures such as moiré systems \cite{PIMENTAMARTINS2023Pressure,ZHAO2021Dynamic, YANKOWITZ2018Dynamic, BRZEZINSKA2024Pressuretuned, XIE2023PressureInduced}.

\begin{acknowledgments}
We thank Y. Wang and L. Xian for helpful discussions. This work was supported by the National Key R\&D Program of China (Grant Nos. 2021YFA1401300 and 2023YFA1406000), the National Natural Science Foundation of China (Grant Nos. 12174439, 12204514, 12174291, and 12274477), the Innovation Program for Quantum Science and Technology (Grant Nos. 2021ZD0302400 and 2021ZD0302300). H. Y. also acknowledges support from the Department of Science and Technology of Guangdong Province in China (2019QN01X061).

X. Z. and J. S. contributed equally to this work.
\end{acknowledgments}

\bibliography{ref.bib}

\end{document}